\documentclass[twocolumn,english,showpacs,superscriptaddress,preprintnumbers,floatfix]{revtex4}
\usepackage[T1]{fontenc}
\usepackage[latin1]{inputenc}
\usepackage{amssymb}

\makeatletter

\providecommand{\LyX}{L\kern-.1667em\lower.25em\hbox{Y}\kern-.125emX\@}
\usepackage{graphicx}
\usepackage{graphicx}

\usepackage{graphicx}

\usepackage{graphicx}

\usepackage{graphicx}

\def\be{\begin{equation}}
\def\ee{\end{equation}}
\def\bea{\begin{eqnarray}}
\def\eea{\end{eqnarray}}

\usepackage{babel}
\makeatother
\begin{document}

\preprint{astro-ph/0411273}

\title{On the Effects due to a Decaying Cosmological Fluctuation}

\author{Luca Amendola}

\email{amendola@mporzio.astro.it}

\affiliation{INAF/Osservatorio Astronomico di Roma, Via Frascati 33, I-00040 Monte
Porzio Catone -- Italy}

\author{Fabio Finelli}

\email{finelli@bo.iasf.cnr.it}

\affiliation{INAF-CNR/IASF, Istituto di Astrofisica Spaziale e Fisica Cosmica,
Sezione di Bologna \\
 via Gobetti, 101 -- I-40129 Bologna -- Italy}

\date{\today{}}

\begin{abstract}
We present the initial conditions for a decaying cosmological perturbation
and study its signatures in the CMB anisotropies and matter power
spectra. An adiabatic decaying mode in presence of components which
are not described as perfect fluids (such as collisionless matter)
decays slower than in a perfect-fluid dominated universe and displays
super-Hubble oscillations. By including a correlated decaying mode with a red
or a scale invariant spectrum, the anisotropy pattern shows super-imposed
oscillations before the first Dopplear peak, while with a blue spectrum
the amplitude of the secondary peaks relative to the first one and
the matter power spectrum can be altered. WMAP first year data 
constrain the decaying to growing ratio of scale invariant adiabatic 
fluctuations at the matter-radiation equality to less than $10 \%$. 
\end{abstract}


\maketitle
\emph{Introduction}. The temperature and polarization anisotropy pattern
of the Cosmic Microwave Background Radiation (CMB) is an invaluable
snapshot of the initial conditions for the primordial perturbations
responsible for structure formation. The most general initial condition
in the linear regime can be written as a sum of several correlated
modes that can be classified as adiabatic or isocurvature depending
on whether the total non-adiabatic pressure perturbation vanishes
or not, and as growing or decaying depending on their time behavior.
The standard paradigm states that the fluctuations begin in their
pure adiabatic mode and that the decaying component becomes rapidly
negligible. Any deviation from this simple one-mode description may
be a signal of new phenomena in the early universe. For instance,
additional isocurvature (growing) modes \cite{bmt1} may be the remnants
of multi-field inflation \cite{bmt1,bmt2,iso-old} or of primordial
phase transitions and a rich literature has developed on this topic.
Although the most recent observations by the WMAP satellite have shown
that the cosmological perturbations start in an initial condition
very close to the growing adiabatic one \cite{iso-wmap}, a sizable
amount of isocurvature perturbations either in the neutrino or in
the CDM component is still allowed by the data \cite{iso-new}.

In contrast, it is rather remarkable that, to the best of our knowledge,
no attempt to constrain the decaying modes on the CMB or matter power
spectrum has ever been performed; in fact, the CMB spectrum of the
decaying mode has never been evaluated. All observational constraints
on the initial conditions neglected to take into consideration the
adiabatic decaying mode, on the implicit assumption that they are
completely negligible. There are however three main reasons why the
decaying modes might be observable on the CMB. First, if they are
generated with a spectrum different from the growing ones, then they
might be dominant either at small or large scales whatever their amplitude
at a given scale is. Second, although the decaying modes die out during
a cosmological epoch, any sharp transition to a different epoch characterized
by different components or different equation of states will regenerate
a new growing and a new decaying modes: the decaying modes trace then
the occurrence of cosmological transitions. Third, if a non standard
era occurred in the early universe (for instance a mix-master singularity
or a bounce \cite{misner}), an adiabatic decaying mode with a large
initial amplitude may be present at the onset of the expanding stage. 
Therefore, it may be conceivable to have a non-vanishing decaying 
part when fluctuations reenter the Hubble radius in bouncing models.
%In particular, in a symmetrical bounce one can expect quite
%generically that decaying and growing modes possess a comparable 
%magnitude at horizon reenter. 
Aim of this paper is to discuss from a general
and phenomenological point of view the general adiabatic mode and
to derive constraints from a likelihood analysis of CMB data. In this
new era of precision cosmology, this is a novel test for the coherence
of primordial cosmological fluctuations.

\emph{The decaying initial conditions}. In order to concretely discuss
this subject it is useful to write the perturbed scalar part of Robertson-Walker
metric both in the longitudinal and in the synchronous gauge, respectively:
\begin{eqnarray}
ds^{2} & = & a^{2}(\tau )[-d\tau ^{2}(1+2\Psi )+dx^{i}dx^{j}\gamma _{ij}(1-2\Phi )]\nonumber \\
 & = & a^{2}(\tau )[-d\tau ^{2}+dx^{i}dx^{j}(\gamma _{ij}+h_{ij})]\, ,
\end{eqnarray}
where $\gamma _{ij}$ is the flat background metric 
and $a(\tau )$ is the scale factor (we fix the present value
$a_{0}=1$ and we have omitted the differences in the coordinates in the 
two gauges in Eq. (1)). The scalar metric perturbation in the synchronous 
gauge
$h_{ij}$ is separated in its trace $h$ and traceless potential $6\eta $.
In a single perfect fluid dominated universe $\Phi =\Psi $, and the
Fourier components of the gravitational potential $\Phi $ obey the
following (driven) dissipative wave equation which admits two linearly
independent solutions \cite{MFB}: \begin{equation}
\Phi ''+3{\mathcal{H}}(1+w)\Phi '-w\nabla ^{2}\Phi +\left[2{\mathcal{H}}'+(1+3w){\mathcal{H}}^{2}\right]\Phi =0\end{equation}
 where $w=p/\rho $ is the equation of state of the fluid and $\mathcal{H}$
is the conformal Hubble parameter. By considering as an example a
radiation dominated universe ($a(\tau )\propto \tau $), $\Phi $
admits as solution: \begin{equation}
\Phi _{k}=\frac{\bar{A}(k)}{y^{3}}\left[\sin (y+\beta _{k})-y\cos (y+\beta _{k})\right]\, ,\end{equation}
 where $y=k\tau /\sqrt{3}$. 
This means that by studying the impact of the decaying mode
we are also testing the coherence hypothesis of the primordial cosmological
fluctuations, which is at the core of the inflationary scenario combined
with a standard cosmology after nucleosynthesis.

Let us now derive the initial conditions at large scales near or after
the epoch of nucleosynthesis, around $z\sim 10^{8}-10^{9}$. At this
stage baryons, CDM, photons are treated as perfect fluids, while neutrinos
\emph{freestream} since they already decoupled from the primordial
plasma (around $z\sim 10^{11}$). We work from now on in synchronous
gauge, as in most Boltzmann codes. The indexes $b,c,r,\nu $ will
denote baryons, CDM, photons and neutrinos, respectively. Without
neutrinos the initial conditions to the lowest non trivial order are,
up to an arbitrary overall factor $C$ \cite{MB}: \begin{eqnarray}
h & = & x^{2}+Dx\nonumber \\
\eta  & = & 2-\frac{1}{18}x^{2}+\frac{3D}{4x}\nonumber \\
 &  & \nonumber \\
\theta _{r} & = & -\frac{k}{18}x^{3}-\frac{3Dk}{8}\label{perfect}
\end{eqnarray}
where $\theta_r$ is the peculiar velocity divergence. These equations are to be complemented by  $\delta _{r}=-2h/3$, $\delta _{c}=\delta _{b}=-h/2$,
$\theta _{c}=0$ and $\theta _{b}=\theta _{r}$, where $D$ is the
amplitude ratio decaying-to-growing and $x=k\tau $ (Note that during
radiation 
$\tau \approx a_{\textrm{eq}}^{-1/2}a/H_{0}$
if $a_{\textrm{eq}}$ is the equivalence scale factor, normalized in such 
way that $a_0=1$ today. In principle this relation receives a small 
correction from the recent $\Lambda$-dominated epoch, but for semplicity 
we shall neglect it thereafter.).

Adding neutrinos is not an easy task, since neutrino traceless pressure
perturbations couple to the shear of the metric ($\Phi \ne \Psi $
in the longitudinal gauge). The neutrino energy-density fraction $R_{\nu }$
is defined as \begin{equation}
R_{\nu }=\frac{\rho _{\nu }}{\rho _{\nu }+\rho _{\gamma }}\quad \quad \quad \frac{\rho _{\nu }}{\rho _{\gamma }}=\frac{7}{8}\left(\frac{4}{11}\right)^{\frac{4}{3}}N_{\nu }\end{equation}
 with $\rho _{\gamma }$, $\rho _{\nu }$,  are the energy
density of neutrinos and photons (after electron-positron annihilation) and $N_{\nu }$ is the number of neutrino species, respectively.
On considering massless neutrinos $R_{\nu }$ is constant in time
(this is also a very good approximation for realistic neutrinos masses).
The physical initial conditions in presence of neutrinos are (again up to an overall factor): \begin{eqnarray}
h & = & \, x^{2}+D\, x^{3/2}\sin X\nonumber \\
\eta  & = & \left[2-\frac{5+4R_{\nu }}{6(15+4R_{\nu })}x^{2}\right]\nonumber \\
 &  & +\frac{D}{x^{1/2}}\left[\frac{11-16R_{\nu }/5}{8}\sin X+5\frac{\gamma }{8}\cos X\right]\nonumber \\
\delta _{\nu } & = & -\frac{2}{3}x^{2}+Dx^{3/2}\left[\left(\frac{1}{4R_{\nu }}-\frac{2}{5}\right)\sin X\right.\nonumber \\
 &  & \left.-\frac{\gamma }{4R_{\nu }}\cos X\right]\nonumber \\
\theta _{\nu } & = & -\frac{23+4R_{\nu }}{18(15+4R_{\nu })}kx^{3}+\frac{D}{16R_{\nu }}kx^{1/2}\times \nonumber \\
 &  & \left[\left(-3-\frac{72}{5}R_{\nu }\right)\sin X+\gamma \left(3-\frac{8}{5}R_{\nu }\right)\cos X\right]\nonumber \\
\theta _{r} & = & -\frac{1}{18}kx^{3}+\frac{D\, k\, x^{5/2}}{3(25+\gamma ^{2})}\left[\gamma \cos X-5\sin X]\right.\nonumber \\
\sigma _{\nu } & = & \frac{4}{3(15+4R_{\nu })}x^{2}+\frac{D}{x^{1/2}}\left[\frac{\gamma }{2}\cos X\right.\nonumber \\
 &  & \left.+\frac{11-16R_{\nu }/5}{10}\sin X\right]\, ,\label{eq:real}
\end{eqnarray}
 where $X=(\gamma /2)\log x+\varphi $ and \begin{equation}
\gamma =\sqrt{\frac{32}{5}R_{\nu }-1}\, .\end{equation}
 To these one must add $\delta _{r}=-2h/3$, $\delta _{c}=\delta _{b}=-h/2$,
$\theta _{c}=0$ and $\theta _{b}=\theta _{r}$ as above. A similar
result was also found by Zakharov \cite{zakharov} for a system of
radiation plus a collisionless component (see also \cite{rebhan}):
our solution include baryons and CDM as well. Notice that $h,\eta$ and $\theta_r$ in (\ref{eq:real})
reduce to (\ref{perfect}) when $R_{\nu }\rightarrow 0$, up to a
trivial rescaling. In the following we will refer to the growing modes
as $g$-modes and to the decaying modes as $d$-modes. We assume that
$C,\, D$ are multiplied by their initial power-law spectrum with
powers $n_{g},\, n_{d}$, respectively, while we take $\varphi $
independent of $k$. This means that the two linearly independent
decaying solutions have the same spectral index.

Several comments are in order. First, the decaying initial condition
(\ref{eq:real}) is as \emph{adiabatic} as one can get: only $\delta _{\nu }$
does not agree with the adiabatic relation $\delta _{\nu }=-2h/3$
since it is the only component which has non-vanishing anisotropic
pressure terms. Second, the decaying mode contains two arbitrary constants,
$D$ and $\varphi $, instead of just one, as for isocurvature modes:
this happens because the order of the system of differential equations
is increased by using a first-order time equation (introduced by the
Boltzmann hierarchy for the moments of the neutrino distribution function)
for $\sigma _{\nu }$ and the number of decaying modes is doubled. By 
considering for simplicity the system of
radiation plus neutrinos, the order of the linear differential equation
for $h$ (see Eq. (93) of \cite{MB}) at leading order in $x$ is
therefore increased from four to five, leading to the growing and
two decaying modes presented here, plus the two gauge modes proper
of the synchronous choice \cite{bmt1}, as can be seen from the master
equation for $h$: 
\begin{eqnarray}
\tau^2 \frac{d^5 h}{d \tau^5} &+& 7 \tau \frac{d^4 h}{d \tau^4} + \left( 5 
+ \frac{8}{5} R_\nu \right) \frac{d^3 h}{d \tau^3} 
\nonumber \\
&+& \frac{24 R_\nu }{5 \tau} 
\frac{d^2 h}{d \tau^2} - \frac{24 R_\nu }{5 \tau^2} \frac{d h}{d 
\tau} = 0
\label{eq:masterh}
\end{eqnarray}
 Third, the mode decays \emph{more slowly} than in a perfect-fluid
dominated universe ($\tau ^{-1/2}$ instead of $\tau ^{-1}$ in Eq.
(\ref{perfect}) in $\eta $) and \emph{oscillates} in $\log x$ with
frequency $\gamma /2$, when $R_{\nu }>5/32$. Oscillations are erased
when $R_{\nu }\leq 5/32$; however this does not occur here since
we keep the conservative value $N_{\nu }=3$. Fourth, the above initial
condition imply that $F_{\nu \, 3}\sim {\mathcal{O}}\left((ka)^{1/2}\times \cos \, ,\sin \right)$.
Therefore the third and the higher moment of the neutrino distribution
are regular in the past ($F_{\nu \, m}\sim {\mathcal{O}}\left((ka)^{m-5/2}\times \cos \, ,\sin \right)$
with $m\geq 3$) and can be safely neglected in the above initial
conditions, as for the growing mode 
\footnote{We have explicitly checked that by adding the correct initial conditions
for $F_{\nu \, 3}$ the resulting power spectrum does not show differences
with respect to $F_{\nu \, 3}=0$ initially.
}.

\emph{Impact on observations}. There are several novel effects due
to the adiabatic decaying mode on the CMB temperature, polarization
and matter power spectra. The oscillations in the gravitational potential
will induce an oscillating pattern of SW and ISW effects at low multipoles
(before the first peak). There is a chance that the waves at $\ell \lesssim 200$
seen in the first year WMAP temperature data are a signature of a
non-vanishing decaying mode on large scales: we are exploring quantitatively
this issue \cite{new}. Note that the superimposed oscillations induced
by a decaying mode appear \emph{only} on scales larger the first acoustic
peak,
% this feature is simply explained by roughly considering $\ell $
% smaller (larger) that the first peak dominated by large (small) scale
% modes, and reminding that 
since the oscillations in $\log x$ found here
are proper only of large scales. The prediction of superimposed oscillations
only before the first peak  sharply distinguishes
the effect of a decaying mode from designing oscillations on the primordial
Fourier matter power spectrum \cite{mr}.

The oscillations are present when $g$ and $d$ modes are correlated:
the pure decaying spectrum, being quadratic in the oscillating functions,
shows in fact attenuated $d$-oscillations. This effect will be maximized
for $n_{d}\leq n_{g}$, because in this case the largest scales are
enhanced with respect to the smaller ones, and also for small optical
thickness $\tau $, because an early reionization erases part of the
oscillation pattern. As is also clear from Fig. 1, the contribution from a 
correlated decaying adiabatic mode is in general not vanishing for $\ell $ larger than
the first peak.

%Remarkably, the contribution from a decaying adiabatic mode is in
%general not vanishing for $\ell $ larger than
%the first peak. 
%The reason
%for this effect is that after horizon crossing the decaying mode
%behaves qualitatively like the growing one.
%Tipically, a decaying mode correlated (anti-correlated)
%with the growing mode enhances (decreases) the power at high $\ell$,
%as can be seen from Figs. (1,2).
%
\begin{figure}
\includegraphics[scale=0.35]{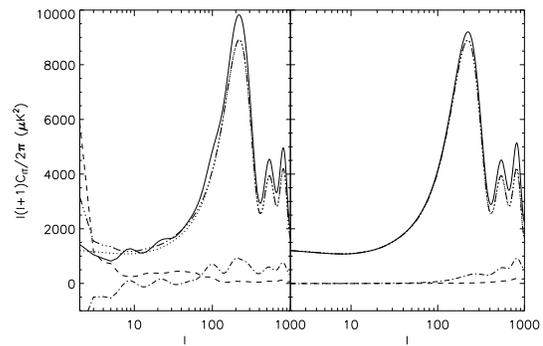}

\caption{Temperature power spectra for \protect{}$D=0.2\, , 
\varphi=\pi/4\, ,n_{d}=1$ (left)
and \protect{}$D=0.2\, , \varphi=0\, ,n_{d}=3$ (right). The other 
parameters are $n_{g}=1$ and $h=0.7,\Omega _{b}h^{2}=0.025,\Omega 
_{c}h^{2}=0.125$.
The solid line is the maximally correlated spectrum, the dotted has
no decaying mode, the 3 dotted-dashed is the uncorrelated spectrum,
the dashed is the decaying spectrum (multiplied by $3$) and the dot-dashed 
the correlation. Note that in the left panel the correlation is negative 
on $\ell \lesssim 60$ and positive on larger $\ell$.} \end{figure}

\begin{figure}
\begin{tabular}{ccc}
 \includegraphics[scale=0.16]{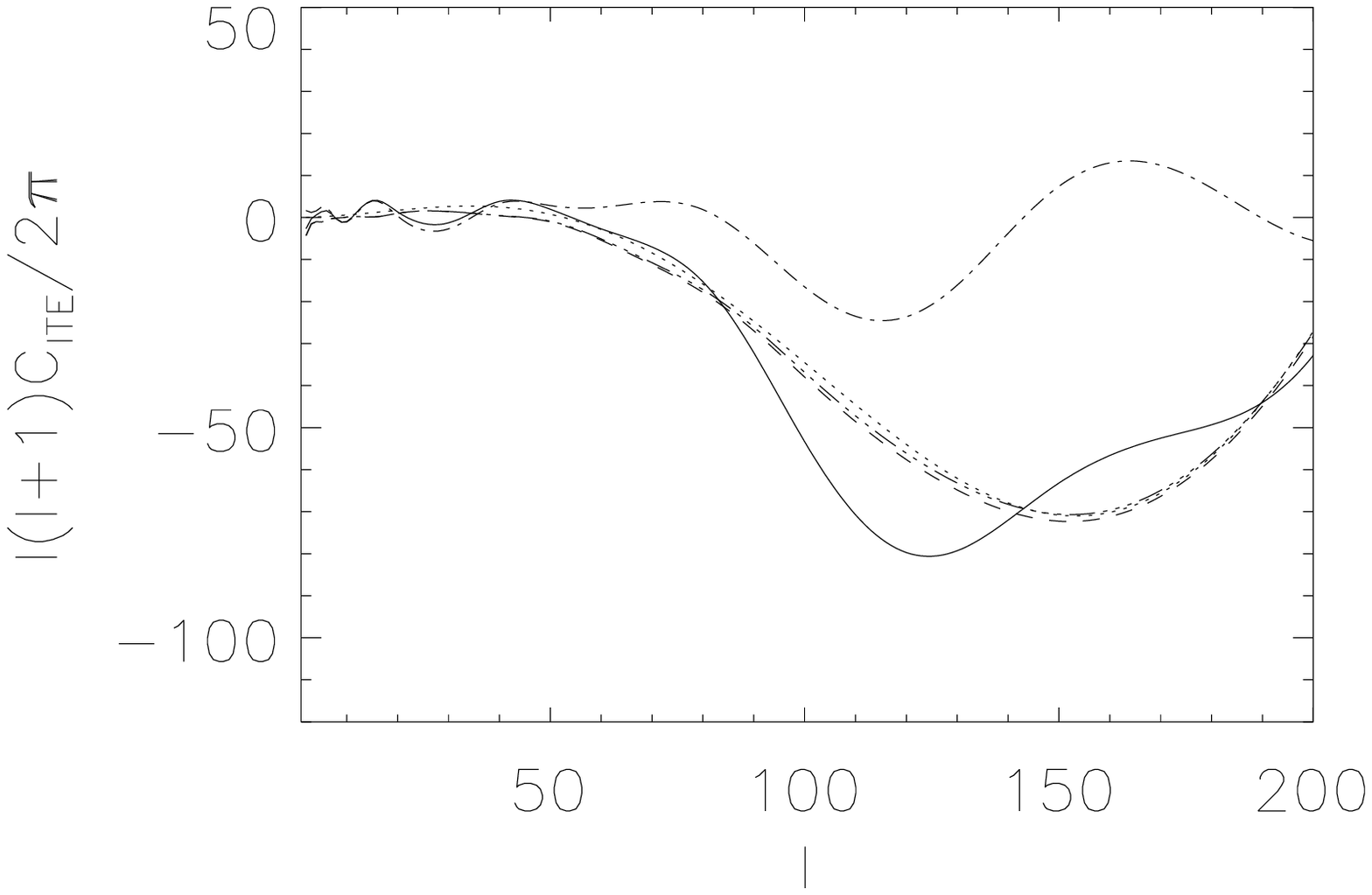}&
 \includegraphics[scale=0.16]{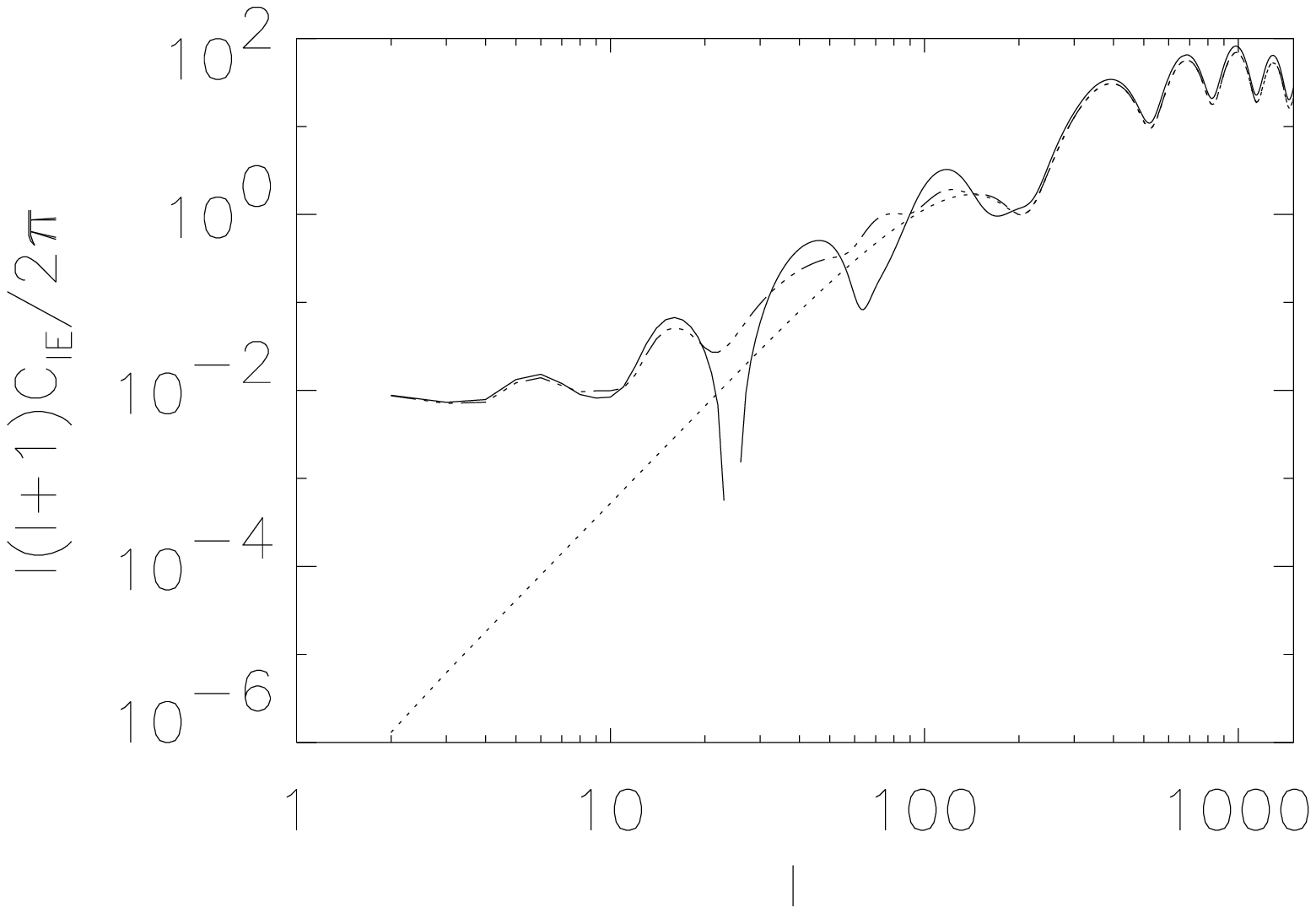}&
 \includegraphics[scale=0.16]{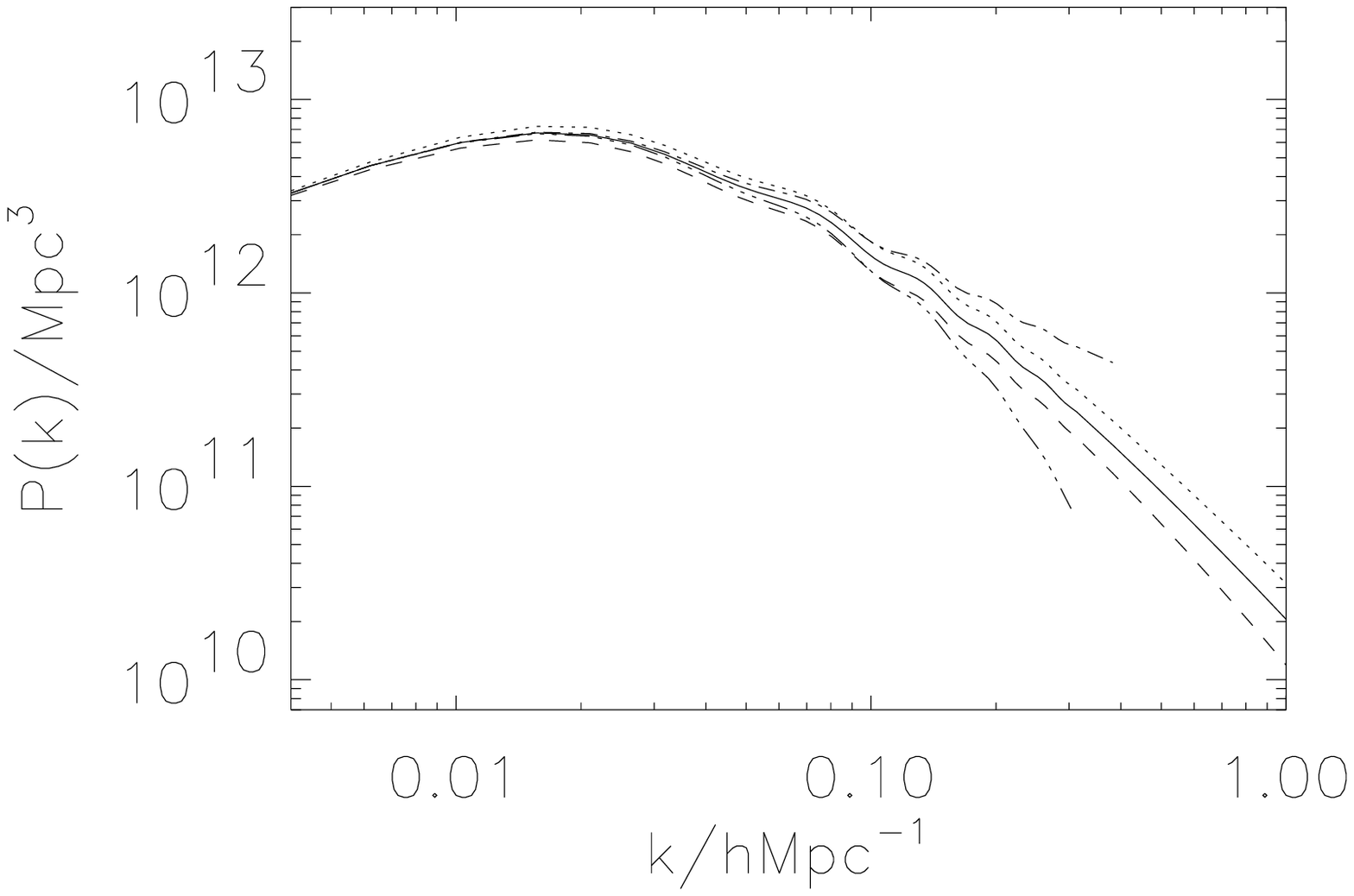}\\
\end{tabular}

\caption{Left and middle panel, TE and E power spectra for the same parameters
of the left panel in Fig. 1 (the convention with lines are the same
as in Fig. 1 and the pure decaying spectrum is not plotted). To the
right, unnormalized power spectrum for CDM for a vanishing decaying
mode (solid line), for \protect{}$n_{d}=1$ \protect{}$D=0.2\, ,-0.2$
(dotted and dashed, respectively), \protect{}$n_{d}=3$, \protect{}$D=0.2\, ,-0.2$
(dot-dashed and 3-dot-dashed).}
\end{figure}

\emph{Likelihood analysis}. Multiplying the decaying mode by the decaying
spectrum $P_{d}\sim k^{n_{d}}$, the fractional amplitude of mode
$k$ in the metric perturbation $h$ can be written as $D_{k}(\tau)
=D(k/k^{*})^{n_{d}/2}(k\tau )^{-1/2}$
where we put $k^{*}=0.05h/$Mpc. We can perform therefore a quick
estimation of the expected effect. For a flat spectrum ($n_{d}=1$)
the fractional amplitude of the decaying mode with respect to the
growing one can be approximated as 
$D_{k^*} (\tau) =D(k^{*}\tau )^{-1/2}\approx D(H_{0}\sqrt{a_{\textrm{eq}}}
/ak^{*})^{1/2}\approx 0.01Da^{-1/2}$.
Extrapolating to the equivalence time ($a_{\textrm{eq}} \sim 3 \times 
10^{-4}$ \cite{wmapdata}) , this amounts to 
$D_{k^*}(a_{\textrm{eq}})\approx 0.56 \, D$.
As from Fig. 1, values of $D_{k^*} (a_{\textrm{eq}}) \sim 
{\mathcal{O}}(10^{-1}-10^{-2})$
(with $n_{d}\approx n_{g}\approx 1$) lead to waves super-imposed
to the power spectrum of a few percent amplitude, similar to those
seen in the first year WMAP data. This also shows that for $n_{d}>n_{g}$
the decaying mode may lead to effects on small scales, as can be seen
from Fig. 2, and can be constrained by {\sc Planck} \cite{planck} and LSS 
data. 
%However,
%this extrapolation clearly fails in the matter regime after equivalence
%so we cannot expect it to be very accurate. 

In a subsequent paper we will investigate in detail the full likelihood
of the model including the features of the low multipole waves \cite{new}.
In this paper we present only the main result: the constraints from
the WMAP data on the decaying amplitude $D$. To simplify the comparison
to real data we assume here that the $g-d$ modes are either maximally
correlated, uncorrelated or maximally anti-correlated . There is no
loss of generality by assuming therefore $D>0$. Inserting the initial
conditions (\ref{eq:real}) in the Boltzmann code produces maximally
correlated spectra; the uncorrelated and anticorrelated cases are
obtained by linear combinations of a basis of three maximally correlated
spectra (for each combination of parameters), as in e.g. \cite{iso-old}.

\begin{figure}
\includegraphics[bb=54bp 500bp 558bp 720bp,clip,scale=0.47]{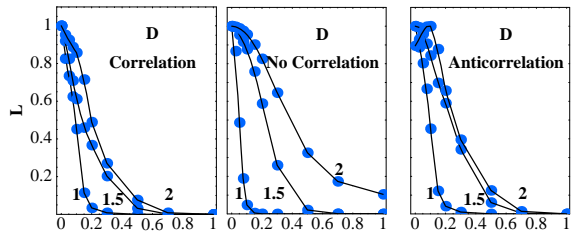}

\caption{Likelihood for the decaying amplitude $D$ for various choices of
$n_{d}$ and of the $g-d$ correlation as in the labels, marginalized
over the other parameters. }
\end{figure}

Values of $n_{d}$ substantially larger than unity generate strong
deviations at small scales, both in the $C_{\ell }$ and in the matter
power spectrum, so they are expected to be severely constrained by
large scale structure data and by large-$\ell $ data \cite{new}.
Fig. 2 shows for instance that the matter power spectrum for $n_{d}>3$
is extremely deficient at small scales, so we exclude such values
from the analysis. We limit ourselves to discuss a few cases: a scale
invariant $n_{d}=1$ and two blue spectra, $n_{d}=1.5,2$. We explored
a grid of over $10^{6}$ models spanning eight parameters (for each
choice of the $d-g$ correlation): $D,n_{d},\varphi $ plus the standard
parameters $n_{s},h,\Omega _{b}h^{2},\Omega _{c}h^{2},\tau $ (the overall
amplitude 
 is automatically integrated out). We restrict ourselves to the massless neutrinos case. We compare
the predictions obtained from a modified version of CMBFAST 
\cite{LOS} with the
first year temperature and polarization WMAP data \cite{wmapdata}.
The global best fit is for $D=1,n_{d}=2,\varphi =-0.785$ rad , no $g-d$ 
correlation, and $n_{s}=0.98,h=0.69,\Omega _{b}h^{2}=0.024$,
$\Omega _{c}h^{2}=0.127,\tau =0.15$, with $\chi ^{2}=970$ (for the
temperature spectrum alone) which is however just marginally better 
($\Delta \chi ^{2}=4$) than the standard $\Lambda $CDM best fit. The 
results of the likelihood
analysis are shown in Figs 3. For a flat spectrum ($n_{d}=1$) the
data constrain $D<0.2$ at 95\% CL both for correlated and anticorrelated
modes and $D<0.1$ for uncorrelated modes. This means that 
any process that excited a decaying mode of the same amplitude
and spectrum of the growing one (i.e. $D_{k^*} (a_*) =1$)
has occurred not later than $a_{*}$, where
$a_{*}=10^{-4}D^{2}<4\cdot 10^{-6}$. For larger $n_{d}$ the
upper limit to $D$ moves to larger values; for $n_{d}=2$, an amplitude
$D$ of order unity is allowed in the uncorrelated case. As it can
be seen from Fig. 3, the effect of the $g-d$ correlation is rather
limited. Similarly, within the range of values of $n_{d}$ we adopted
here the phase $\varphi $ has a very small effect on the likelihood
(a larger effect is obtained for $n_{d}<1$). Finally, the estimation
of the other cosmological parameters is not much affected by the addition
of $d$-modes \cite{new}. Therefore, it is safe to consider $D<1$
as an upper limit independent of all the other parameters involved
in the model. Fixing $n_{d}=1$ the upper limit strenghtens to $D<0.2$.

\emph{Conclusions}. We have derived the initial conditions for an
adiabatic decaying mode in a standard primordial soup (neutrinos,
baryons, CDM and radiation) in a spatially flat universe in Eq. 
(\ref{eq:real}). 
We have shown that decaying modes induce super-Hubble oscillations and 
observable deviations at small scales in the CMB and matter power spectra, 
depending on their amplitude and spectrum. WMAP first year data alone 
constrain $D<0.2$ at 95\% CL for a scale invariant spectrum almost 
regardless of the other parameters. This means that scale invariant 
adiabatic fluctuations should  consist of their growing part for at least 
$\sim 90 \%$ at matter-radiation equality. 
When $n_d > n_g$, larger values 
for $D$ are still allowed by observations.  

\vspace{.3cm}

\textbf{Acknowledgements}

We thank the staff at CINECA where most of the computations have been
performed under the grant INAF@CINECA.

\end{document}